\documentclass[useAMS,usenatbib]{mn2e}

\usepackage{graphicx}
\usepackage{amsmath}
\usepackage{natbib}
\usepackage{myjournals}
\begin{document}

\title{Cosmological Simulations of Dwarf Galaxies with Cosmic Ray Feedback}

\author[Chen \& Bryan]{Jingjing Chen$^{1}$, Greg L. Bryan$^{1}$, Munier Salem$^{1}$ \\
Department of Astronomy, Columbia University, New York, NY 10027, USA}

\date{}

\maketitle


\begin{abstract}
We perform zoom-in cosmological simulations of a suite of dwarf galaxies, examining the impact of cosmic-rays generated by supernovae, including the effect of diffusion.  We first look at the effect of varying the uncertain cosmic ray parameters by repeatedly simulating a single galaxy.  Then we fix the comic ray model and simulate five dwarf systems with virial masses range from 8 to $30 \times 10^{10}$ M$_\odot$.   We find that including cosmic ray feedback (with diffusion) consistently leads to disk dominated systems with relatively flat rotation curves and  constant star formation rates.   In contrast, our purely thermal feedback case results in a hot stellar system and bursty star formation.  The CR simulations very well match the observed baryonic Tully-Fisher relation, but have a lower gas fraction than in real systems. We also find that the dark matter cores of the CR feedback galaxies are cuspy, while the purely thermal feedback case results in a substantial core.
\end{abstract}


\section{Introduction} \label{sec:intro}

Isolated dwarf galaxies are commonly observed in the local Universe, and provide an interesting way to test models of galaxy formation and evolution.  Below a stellar mass limit of about $10^{10}$ M$_\odot$ (or rotation velocity of about 100 km/s), such systems are largely star-forming systems except when they become satellites of larger halos  \citep{Geha2012, Bradford2015}.   Naively, one might expect such systems to be easy to understand, as they have small or non-existent bulges and are free from uncertainties related to AGN feedback.  However, detailed comparisons with cosmological models have encountered two kinds of problems.  

The first problem has to do with the baryonic content of the halos -- simply put, galaxy formation must become increasingly less efficient as the halo mass decreases, for standard $\Lambda$CDM to match observations.  It has long been recognized that this is implied by the different slopes of the low-mass end of the dark matter halo mass function and the faint end of the luminosity function \citep{WhiteFrenk1991}.  More recent work has sharpened this mismatch with both semi-analytic models \citep[e.g.,][]{2010Guo, 2013Behroozi, 2012Leauthaud} and simulations \citep[e.g.,][]{2007ApJ...667..170S, 2010Governato, 2010MNRAS.402.1599S}.  A closely related problem is the "missing" dwarf satellites around Milky-Way mass halos \citep{1999ApJ...524L..19M,2008MNRAS.391.1685S}.

A second issue has to do with the inner structure of dwarf galaxies -- in particular the relation between the halo mass and the rotational velocity near the core.  Dark matter only simulations predict quite cuspy profiles with relatively high dark matter densities in the center, while observations generally find cored profiles \citep[e.g.,][]{1994Moore, 2000Salucci, 2002deBlok, 2008Walter}.  A related version of this problem has recently been highlighted among the dwarf satellites as the "too big to fail" problem, which highlights the paucity of satellites with rotation velocities in the 30-70 km/s range \citep{2012Boylan-Kolchin}.

Solutions to both of these problems in an astrophysical context require efficient feedback mechanisms that can blow out a large amount of gas, suppress star formation, bring down the baryonic fraction, and possibly reduce the central dark matter density of lower mass dwarfs.  It has long been clear that the low potential depth of dwarf systems makes them particularly susceptible to stellar feedback \citep{1986Dekel}; however, modeling feedback from supernovae is numerically challenging \citep[e.g.,][]{1994Navarro}.

A variety of numerical feedback mechanisms have been tested in different papers.   For example, by suppressing cooling immediately after a star formation event to counteract the artificially enhanced radiative losses \citep{2007MNRAS.374.1479G, 2011ApJ...742...76G}, simulated galaxies produced less massive bulges.  Supernovae feedback which injects momentum directly in to the surrounding gas has also been shown to drive mass-loaded outflows \citep[e.g.,][]{2008MNRAS.387..577O}.  This has been combined with other physical processes -- for example, \citet{2011MNRAS.413..659S} combined supernovae feedback with photoionization from ultraviolet background, and produced dwarf galaxies with high mass-to-light ratios.

More recent simulations have made progresses in reproducing the observed properties using subgrid models with strong feedback \citep[e.g.,][] {2014Dicintio, 2014Hopkins, 2015Schaye}.  For example, \citet{2015MNRAS.447.3548S} showed that they could reproduce the star formation main sequence relation at $z = 0$.  Zoom-in simulations afford more resolution, and have shown much recent promise in reproducing the stellar mass content of dwarf satellites \citep[e.g.,][] {2010Governato, 2012Governato, 2014Hopkins, 2015Onorbe}.  \citet{2014ApJ...786...87B} found that energetic supernovae feedback and subsequent tidal stripping has a significant effect on reducing the dense core of Milky Way mass galaxies

Here we explore the impact of a specific physical feedback mechanism on dwarf galaxies, namely dynamical cosmic ray (CR) feedback.  CRs are high-energy, quickly diffusing particles accelerated by shock waves.  They have a wide range of energies, with the largest contribution to the energy density coming from those particles with energies around a few GeV.  The energy density of CRs is comparable to the magnetic energy and turbulent energy in galaxies \citep{2005ppfa.book.....K}.

In this paper, we present our cosmological simulations of dwarf galaxies with CR feedback.  We use the {\sc Enzo} code, with an implementation of CR feedback by \citet{2014SB}.  This adopts a relatively simple two-fluid model that provides an extra pressure term and also allows for diffusion of the CRs \citep{1985cgd..conf..131D}.  In \citet{2014SB} and \citet{2016SBC}, we applied this model to the cosmological simulation of a Milky Way mass galaxy.  Here we examine the effect of CR feedback and diffusion on dwarf galaxies with halo masses in the range of 8 to $30 \times 10^{10}$ M$_\odot$.

In section~\ref{sec:method}, we describe the simulation model, the feedback scheme, initial conditions, and the selection of halos for further simulations.  In section~\ref{sec:vary_cr}, we choose the best set of parameters by comparing the simulation result of a single system with varying CR physics.  Then we apply this model to a series of dwarf galaxies, with halo mass spanning from 8 to $30 \times 10^{10}$ M$_\odot$ in section~\ref{sec:vary_mass}, and we evaluate the results by comparing to observed scaling relations.
Finally, in Section~\ref{sec:disc}, we discuss our results.
\\


\begin{figure*}
\centering
  \includegraphics[width=\textwidth]{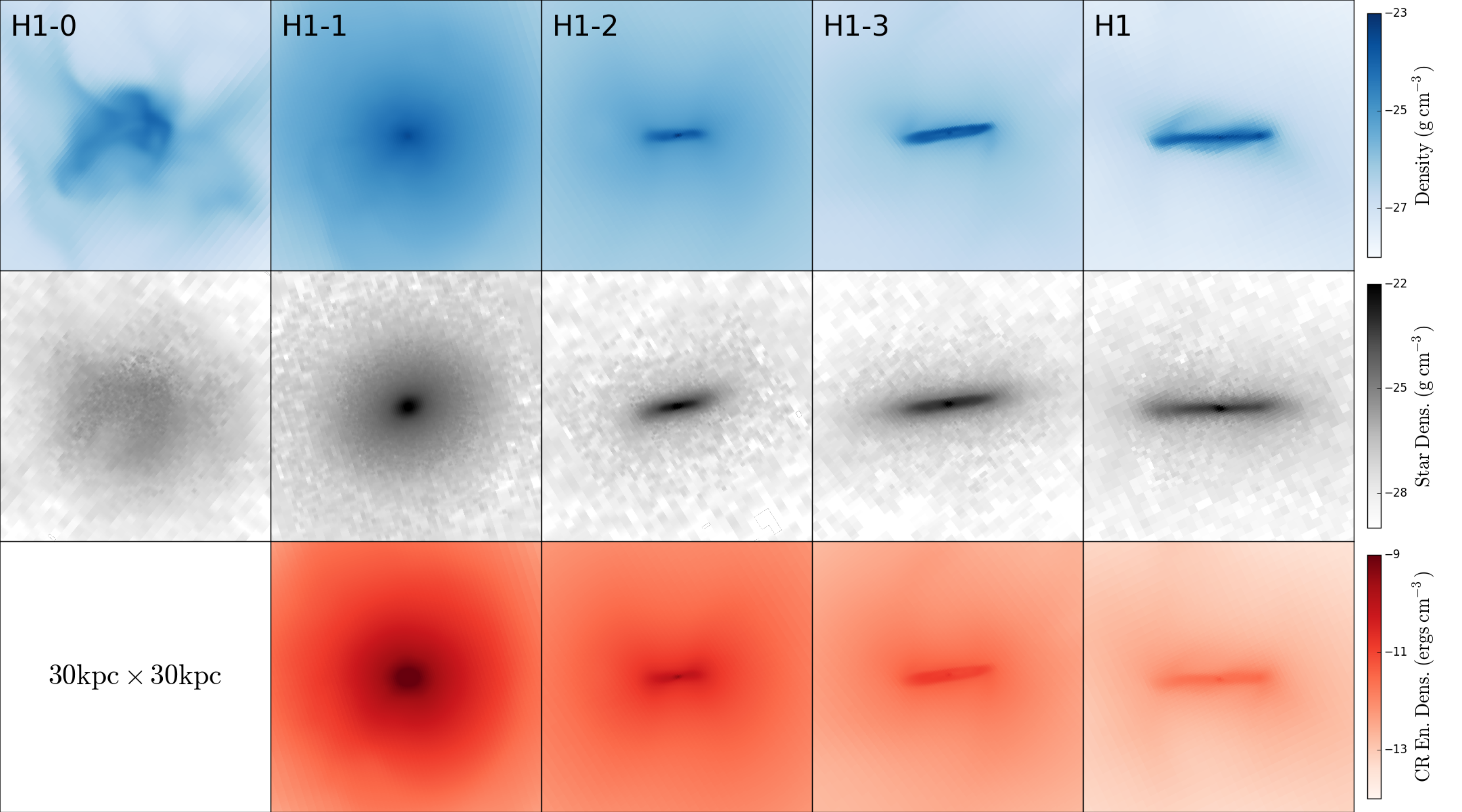}
  \caption{Edge-on views of the same halo modeled with different CR physics.  From left to right, the panels show  a simulation with no CRs, followed by  runs with CR diffusion coefficients of 0, $3\times10^{27}$, $1\times10^{28}$, $3\times10^{28}$ cm$^2$s$^{-1}$. From top to bottom, the panels show the projected (density-weighted) gas density, stellar density, and CR energy density. Each panel represents a $30$ kpc $\times 30$ kpc area.}
  \label{cr_edge}
\end{figure*}


\section{Method} \label{sec:method}

We begin with a brief introduction to our hydrodynamics method, and then describe the two-fluid model used to model the CRs, and finally outline the cosmological initial conditions. 

\subsection{Numerical code: {\sc Enzo}}

Our simulations use the open-source code {\sc Enzo}, which is a three-dimensional, Cartesian, grid-based hydrodynamics code that includes adaptive mesh refinement \citep{1997astro.ph.10187B, 2004astro.ph..3044O, 2012ApJ...749..140H, 2014ApJS..211...19B}. 
We include a non-equilibrium chemical model following H, H$^+$, He, He$^+$, He$^{++}$, and include radiative cooling from all the above chemical species as well as metal cooling computed in a lookup table as described in \citet{Smith2008}.  We include a metagalactic radiative background as computed in \citet{2012Haardt}.  The star formation recipe of {\sc Enzo} is based on the criteria presented in \citet{1992ApJ...399L.113C} and described in detail in \citet{2014ApJS..211...19B}.   Briefly: when a parcel of dense gas is identified to be Jeans unstable and collapsing, some of its gas may be converted into a star particle with some efficiency.   For a more detailed discussion of the star formation (and thermal-only) feedback scheme used in this paper, see \citet{2012ApJ...749..140H} -- in this work we adopt an efficiency of star formation per free fall time of 1\% and an energetic feedback efficiency of $e_{\rm SN} = 3 \times 10^{-6}$ of the stellar rest mass energy.

\subsection{Cosmic Ray Feedback Scheme}

We use a model for CR feedback and diffusion as implemented, tested and described in \citet{2014SB}.  The model uses a two-fluid method \citep{1985cgd..conf..131D, 1986MNRAS.223..353D, 1994ApJ...429..748J}.  It treats the high-energy particles as an ideal gas with $\gamma = 4/3$.  Anisotropic CR pressure and the dynamical impact of magnetic fields are assumed to be subdominant.   The fluid equations for the thermal gas stay as usual, except that the momentum of the gas is affected by the pressure of both the gas and CR components.  In addition, we optionally include isotropic diffusion of the cosmic rays.  Here we simply highlight the additional equation that describes the evolution of the CR energy density $\epsilon_\mathrm{CR}$:
\begin{equation}
\begin{split}
\partial_t \epsilon_\mathrm{CR} + \nabla \cdot (\epsilon_\mathrm{CR} \textbf{\textit{u}}) = 
& - P_\mathrm{CR} (\nabla \cdot \textbf{\textit{u}})  \\
& + \nabla \cdot (\kappa_\mathrm{CR} \nabla \epsilon_\mathrm{CR}) + \Gamma_\mathrm{CR}
\end{split}
\end{equation}
where $\textbf{\textit{u}}$ is the fluid velocity, P$_\mathrm{CR}$ is the pressure from CR, $\kappa_\mathrm{CR}$ is the CR isotropic diffusion coefficient, and $\Gamma_\mathrm{CR}$ is the source term for CR.  Except for the isotropic diffusion, the CRs are assumed to be tightly coupled to the gas through a tangled magnetic field (which is not explicitly simulated).  While this model is approximate, simulations including magnetic fields and/or anisotropic diffusion have also been shown to drive winds with similar mass-loading factors \citep[e.g.,][] {2013Hanasz, 2016Girichidis}.  We also do not include any non-adiabatic losses in the CRs; in reality, CRs lose energy primarily through collisions with the thermal plasma; this is generally subdominant to adiabatic cooling in our simulations except for runs with zero (or low) diffusion coefficients, which are not realistic in any case.  A more detailed discussion of this point and the shortcomings of the model in general can be found in \citet{2014SB} and \citet{2016SBC}.

When a Type II SNe is formed, it will eject some of its mass and energy back into the surrounding cells.  In our model, we assume that some of the energy is in thermal form and some accelerated by blast waves (generally not resolved in the grid).  The fraction of energy that is deposited to thermal gas and CR is 0.7 and 0.3 respectively; this CR fraction is somewhat larger than canonical values but within suggested ranges \citep[e.g.,][]{Wefel1987, Ellison2010, Kang2006}.  Moreover, we assume a relatively steep initial mass function (our chosen $e_{\rm SN}$ corresponds to 1 SN per 185 M$_\odot$ masses of stars produced), and hence a relatively low SN energy injection rate, so the CR energy produced per solar mass of stars formed is within typical estimates.  We adopt these values for consistency with previous work \citep{2012ApJ...749..140H, 2014SB}, where we also explored the impact of varying them.

\subsection{Initial Conditions and Halo Selection}

We are interested in modeling a number of dwarf galaxies forming out of cosmological initial conditions.  Therefore, we use {\sc MUSIC} \citep{2011MNRAS.415.2101H} to set up the initial density and velocities fields.  This is done using the cosmological parameters from the best fit Planck 2013 results \citep{2014A&A...571A..16P}, i.e. 
H$_0$ = 67.11 km/s/Mpc,
$\Omega_\Lambda = 0.6825 $,
$\Omega_\mathrm{m} = 0.3175 $,
$\Omega_\mathrm{b} = 0.0490 $
${\sigma_8 = 0.8344}$. 
All the simulations start at $z =100$ and stop at the present day.

\begin{table}
\centering
\caption{Simulation parameters for fixed halo, varying CR runs.}
\begin{tabular}{c|c|c|c|c}
\hline
ID & CR physics & R$_\mathrm{virial}$ & M$_\mathrm{virial}$ & M$_\mathrm{star}$\\
\hline
H1-0 & no CR 			& $170.13$ & $2.67\times10^{11}$ & $0.15\times10^{10}$   \\ 
H1-1 & $0$ 			& $182.09$ & $3.27\times10^{11}$ & $1.93\times10^{10}$ \\ 
H1-2 & $3\times10^{27}$  & $179.79$ & $3.15\times10^{11}$ & $0.88\times10^{10}$  \\
H1-3 & $1\times10^{28}$  & $177.05$ & $3.01\times10^{11}$ & $1.02\times10^{10}$   \\ 
H1    & $3\times10^{28}$ 	& $174.32$ & $2.87\times10^{11}$ & $1.04\times10^{10}$   \\
\hline
\end{tabular}
\smallskip
R$_\mathrm{virial}$ is in unit of kpc. M$_\mathrm{virial}$ and M$_\mathrm{star}$ are in unit of M$_\odot$. 
\label{diff_cr}
\end{table}

First we perform a dark matter only run in a (20 Mpc/h)$^3$ box and identify halos that are around $10^{10}$ to $10^{11}$ M$_\odot$.  All of these halos become a little more massive in the final runs, listed in Table~\ref{halo_list}. 
To study the impact of CR feedback, we select relatively isolated halos so that the physical properties of the dwarfs are not affected by nearby massive galaxies.  In particular, all the selected systems are at least 0.5 Mpc away from any halos that are more massive than 0.1 times of the target halo.  Then we trace the dark matter particles currently in the halo to their positions at the beginning of the simulation. 
The selected halos are re-simulated at high-resolution with baryons and adaptive mesh refinement (AMR) is performed in the cubical region that encloses all the dark matter particles from the beginning.  Our root grid is 128$^3$ and we use three additional levels in the initial conditions for a particle mass of M$_{DM} = 8.3\times10^5$ M$_\odot$.  During the simulation, additional refinement is added whenever the gas (dark matter) mass in a cell exceeds $6.0\times10^5$ ($3.3\times10^6$) M$_\odot$.  We allow up to 10 levels of refinement, resulting in a smallest cell size of 227 pc.

In order to explore the impact of some of the uncertain parameters in our model, simulations with different CR physics, including a no CR run, and CR runs with diffusion coefficients of $0$, $3\times10^{27}$, $1\times10^{28}$, and $3\times10^{28}$ cm$^2$ s$^{-1}$ are done on the most massive of the selected halos. The results of these simulations are listed in Table~\ref{diff_cr}.  
We analyze the results from these runs (see below) and choose the coefficient of $3\times10^{28}$ cm$^2$ s$^{-1}$ to carry on the simulations with the other selected halos.  As discussed below, this choice is somewhat arbitrary, but it is consistent with estimates from recent GALPROP models \citep[e.g.,][]{Ptuskin2006, Ackermann2012} and with observational measurements \citep[e.g.,][]{Strong1998, Tabatabaei2013}.  It is also found to be consistent with galactic gamma ray emission in \citet{2016SBC}.
The properties of all selected halos simulated with this choice of parameters are listed in Table~\ref{halo_list}.  

\begin{table}
\centering
\caption{Simulated dwarf galaxy properties}
\begin{tabular}{c|c|c|c}
\hline
ID & R$_\mathrm{virial}$ & M$_\mathrm{virial}$ & M$_\mathrm{star}$ \\
\hline
H1  & 174.32 & $2.87\times10^{11}$ & $1.04\times10^{10}$   \\
H2  & 161.82 & $2.30\times10^{11}$ & $7.85\times10^{9}$   \\
H3  & 132.10 & $1.25\times10^{11}$ & $2.98\times10^{9}$  \\
H4  & 115.78 & $8.40\times10^{10}$ & $1.52\times10^{9}$   \\
H5  & 118.60 & $9.05\times10^{10}$ & $1.25\times10^{9}$   \\
\hline
\end{tabular}
\\
\smallskip
R$_\mathrm{virial}$ is in unit of kpc. M$_\mathrm{virial}$ and M$_\mathrm{star}$ are in unit of M$_\odot$. 
\label{halo_list}
\end{table}


\section{Results} \label{sec:result}

First, we focus on comparisons of runs with different CR physics, using the most massive of the halos.
Second, we discuss the physical properties of the suite of five halos with fixed CR physics.
Third, we compare our simulation results to observations, specifically the Baryonic Tully-Fisher relation and HI stellar mass relation.  Analysis has been carried out using the {\it yt} package \citep{2011Turk}.


\subsection{Impact of Varying CR Physics}
\label{sec:vary_cr}

\begin{figure}
\centering
  \includegraphics[scale=0.45]{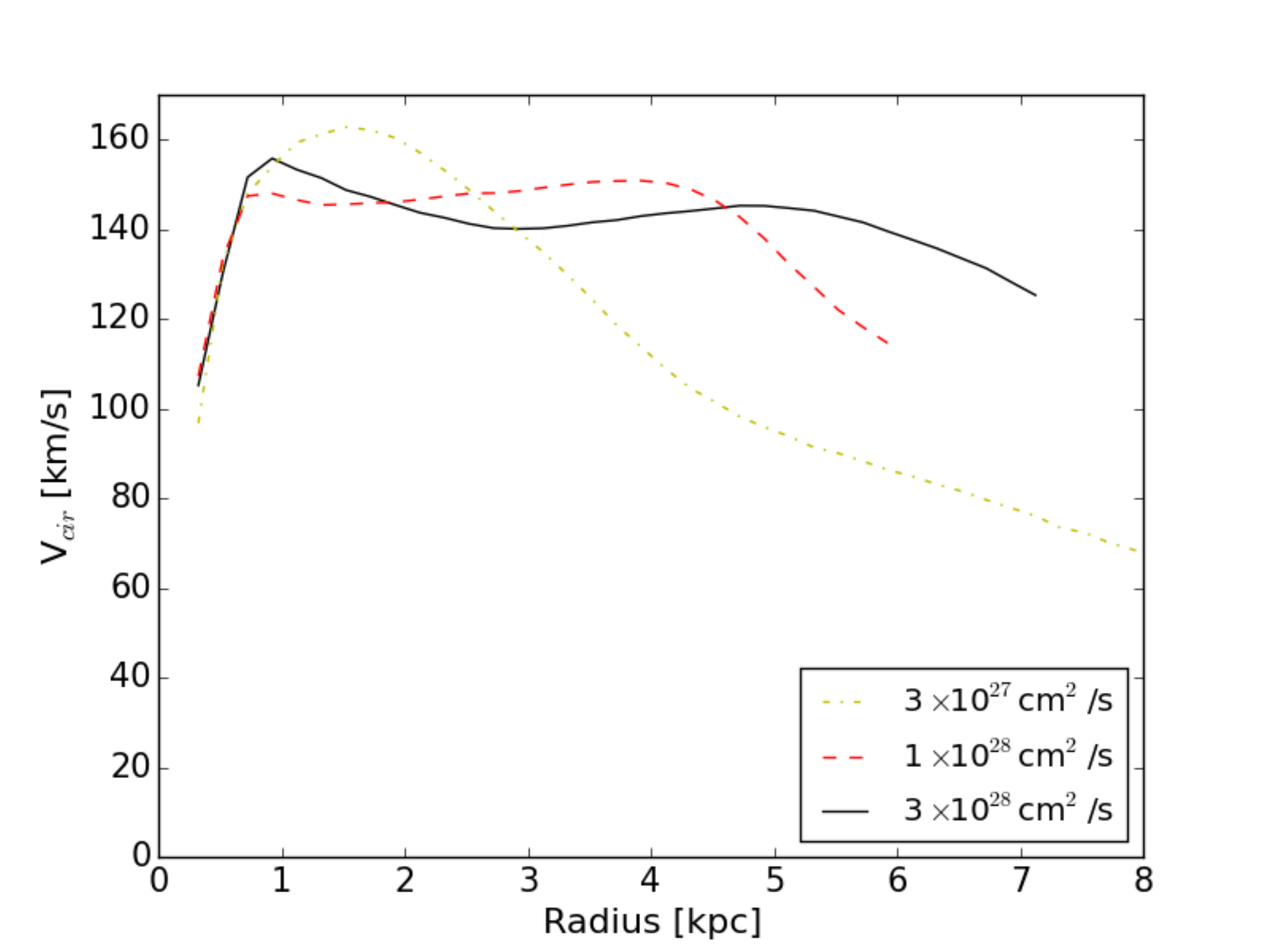}
  \caption{Circular velocity profiles of the neutral hydrogen gas in the disk from the simulations with non-zero CR diffusion, measured out to a limiting HI surface density of $10^{20}$ cm$^{-2}$.}
  \label{cr_vcir}
\end{figure}

We choose an isolated halo of around $3 \times10^{11}$ M$_\odot$ and study the effect of CR feedback by varying the CR physics and fixing all the other parameters (See Table~\ref{diff_cr} for the simulation parameters).
Figure~\ref{cr_edge} shows edge-on images of the gas density, stellar density and CR energy density of the dwarf galaxies produced in each run at $z=0$, focusing on just the center of the halo where the stellar system forms.  Since H1-0 and H1-1 are not disk-shaped, we show them in the direction of their total angular momentum within 10 kpc radius.

The run without any CR feedback (H1-0; left-most panels) forms a diffuse, irregular cloud.  We find that the energetic thermal feedback is actually remarkably effective in regulating the formation of dense, star-forming clumps.  This behavior is quite different than that seen in high-mass galaxies using identical star formation and feedback prescriptions \citep[e.g.,][]{2012ApJ...749..140H} and is more like the behavior observed in very low-mass galaxies \citet[e.g.,][]{2013MNRAS.432.1989S, 2015ApJ...809...69S}.  We explore this point in more detail in Section~\ref{sec:feedback_eff}

Moving on to the simulations with cosmic-rays, we see in Figure~\ref{cr_edge} that H1-1, with no CR diffusion, has an almost spherically symmetric gas distribution.  This occurs because when the CRs are tied to the gas, they provide complete pressure support and rotational support is unimportant.  Clearly this is unrealistic (and probably violates gamma-ray emission from dwarf irregulars -- see \citet{2016SBC}).

\begin{figure}
\centering
  \includegraphics[scale=0.45]{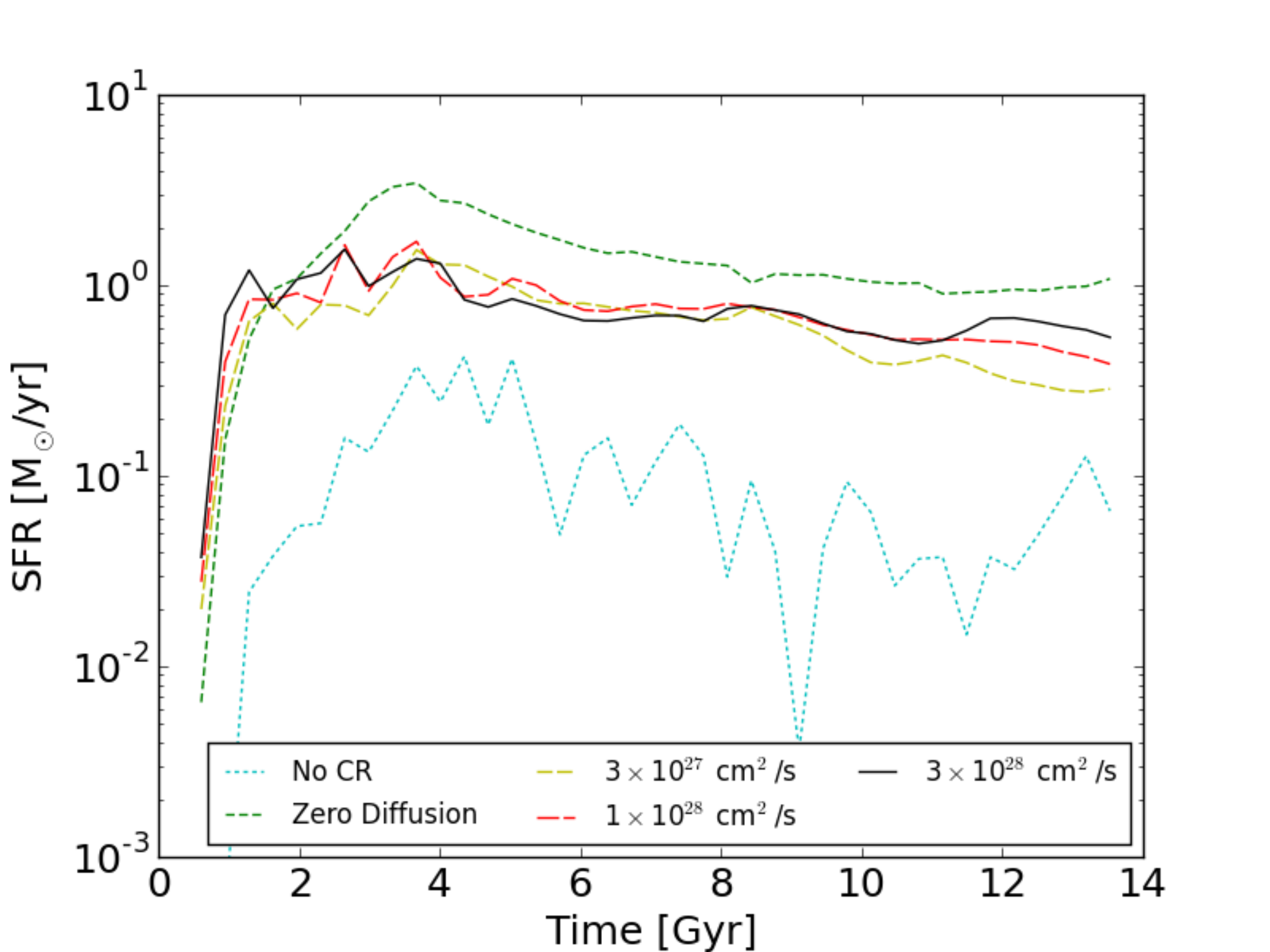}
  \caption{Comparison of the star formation rate history in simulations of the same halo but different CR physics, as noted in the legend. }
  \label{cr_sfr}
\end{figure}

The other runs, which have non-zero CR diffusion, all produce disk-shaped galaxies, generally becoming thinner as the diffusion coefficient increases.  H1, which has the highest diffusion coefficient of $3\times10^{28}$ cm$^2$ s$^{-1}$, results in the most extended disk.  H1-1, which has the lowest non-zero diffusion coefficient, results in a smaller disk which is surrounded by denser gas.  The gas density and CR energy density immediately surrounding the disk also decreases as $\kappa$ increases.  The CRs escape more easily, both because they diffuse out more rapidly and because they drive stronger outflows \citep{2014SB}, and therefore they provide less pressure support, decreasing the weight of gas they can support in the diffuse (not-rotating) halo.

\begin{figure*}
\centering
  \includegraphics[width=\textwidth]{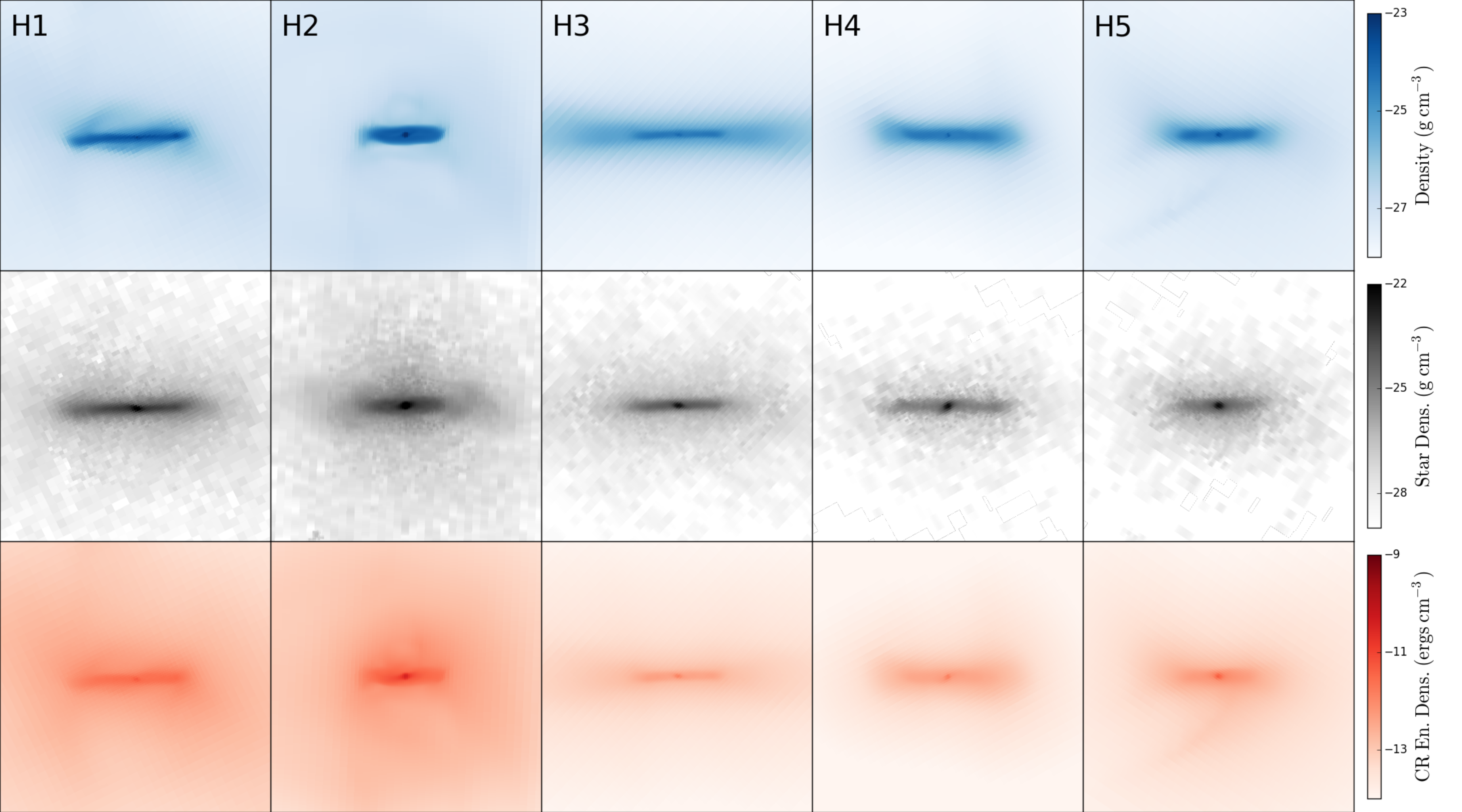}
  \caption{Edge-on views of the simulated galaxies with different halo masses but a fixed diffusion coefficient ($\kappa = 3 \times 10^{28}$ cm$^2$ s$^{-1}$. The plot show halos H1, H2, H3, H4, H5 from left to right, and gas density, stellar density, and CR energy density from top to bottom. Each panel represents a $30$ kpc $\times 30$ kpc area at $z=0$.}
  \label{mass_edge}
\end{figure*}

Since all the runs with non-zero CR diffusion produce disk galaxies, we can examine the rotational velocity of the gas in the disk.  Figure~\ref{cr_vcir} shows the rotational velocities of the HI gas in H1-2, H1-3, and H1.  To compute these rotation velocity curves, we look at the actual rotational velocities of the HI gas (rather than a simple dynamical measure such as $(GM/r)^{1/2}$).  To do this, we adopt $10^{20}$ cm$^{-2}$ as the lowest detectable HI surface density in order to make an approximate comparison to observations -- the results are not strongly sensitive to this threshold, but disks from runs with low diffusion coefficients start to pick up some CR-pressure supported gas and show falling rotation curves.  In particular, in H1-2, the run with a diffusion coefficient of 3 $\times$10$^{27}$ cm$^2$ s$^{-1}$ is affected by this issue: the circular velocity peak at 160 km/s at about 1.5 kpc radius and falls quickly beyond that radius.  As can be seen from its edge-on plot, the disk in H1-2 has a small radius, but its surrounding gas is relatively dense, which makes the HI detectable radius even larger than the disk radius.  In that run, outside the disk, the gas is actually supported in part by the pressure of the CRs, which explains why the circular velocity curve drops so quickly after the peak.  

From this figure, we see that, at least for the two higher diffusion coefficient runs, the circular velocity curves are relatively flat, in reasonable agreement with observed systems.  For example, in the run with the highest diffusion coefficient (H1), the circular velocity is relatively flat at about 140 km/s out to 6-7 kpc.

Next, we examine the impact of CR pressure on the star formation rate.  This is shown in Figure~\ref{cr_sfr} for all the runs of halo H1 with different CR physics.  For all of our more realistic runs, with non-zero diffusion coefficient, the SFRs are quite similar: an early rapid rise follow by a nearly constant rate that falls slowly from 1 to 0.5 M$_\odot$/yr.  The run without diffusion shows a rate which is a factor of two larger, due to the higher gas densities in its core, while the thermal-only feedback is lower by nearly an order of magnitude and shows considerably more variation with time -- consistent with the bursty feedback seen in the gas distribution.


\subsection{Varying Halo Mass with Fixed CR Physics}
\label{sec:vary_mass}

In this section, we fix the CR diffusion coefficient to be $3\times10^{28}$ cm$^2$s$^{-1}$ and study the impact on dwarf galaxies with different halo masses, varying the total mass from about $8 \times 10^{10}$ M$_\odot $ to $3 \times 10^{11}$ M$_\odot $.   We examine how CR feedback behaves in different scales and also use this suite of runs to compare with the observational scaling relations in Section~\ref{comparison_observation}.

All of these runs produce dwarf galaxies with extended disk features; Figure~\ref{mass_edge} shows edge-on images of these galaxies.  The halo masses decrease from left to right: H1 to H5 as described in Table~\ref{halo_list}.  There is a fairly clear trend of decreasing disk extent in the stellar distribution, with the larger halos having more extended disks.  H2 is a slight outlier in this trend with a truncated disk, particularly in the gas and CR components.  In addition, the circumgalactic medium is less dense (with a lower CR energy density) for the lower mass halos.

\begin{figure}
\centering
  \includegraphics[scale=0.45]{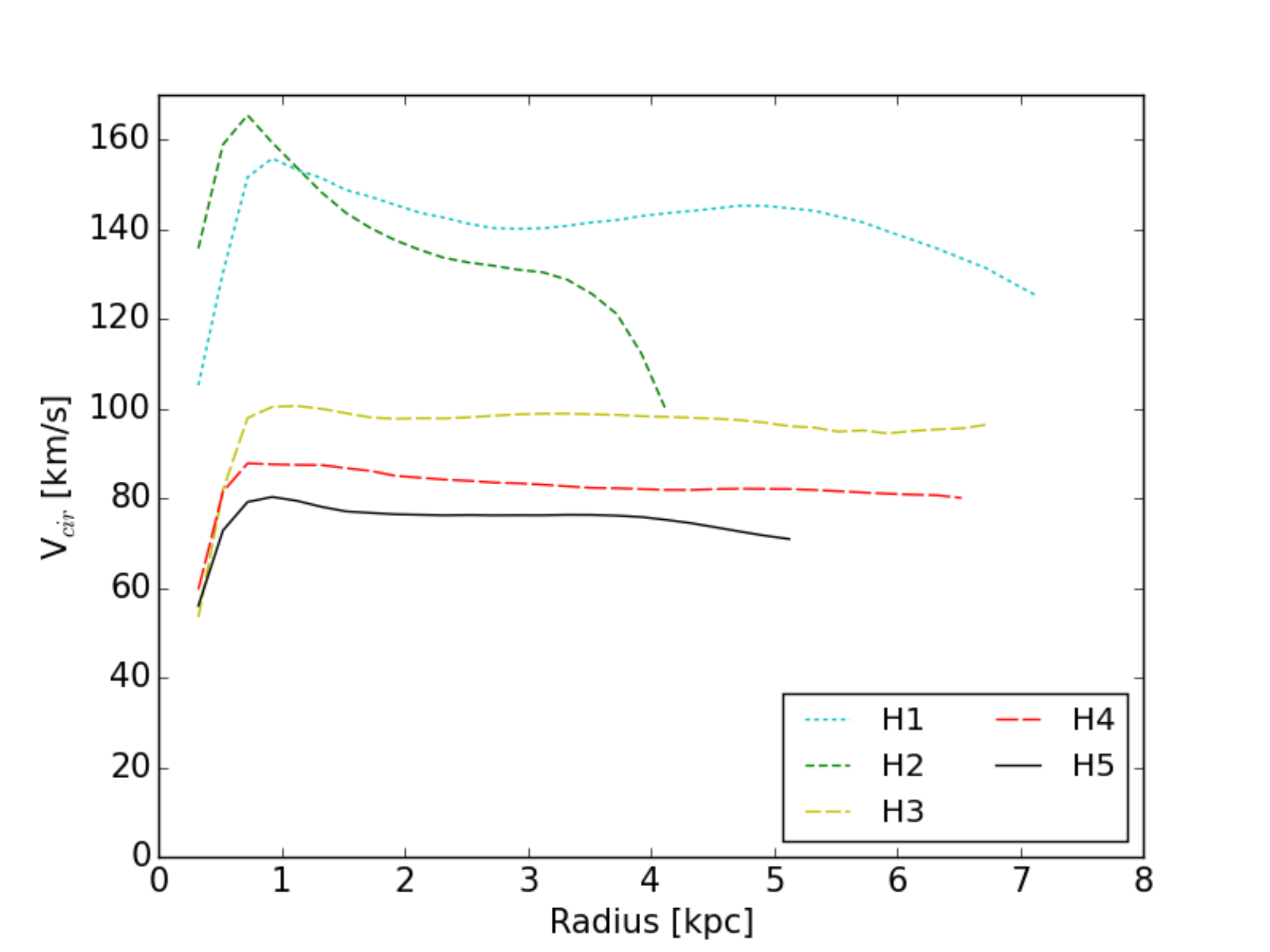}
  \caption{Measured circular velocity of the neutral hydrogen in the disk for halos H1 to H5.  As in Figure~\ref{cr_vcir}, we plot the circular velocity only out to a limiting HI surface density of $10^{20}$ cm$^{-2}$.}
  \label{mass_vcir}
\end{figure}

We plot the rotational velocities of the gas for these five galaxies in Figure~\ref{mass_vcir}.  Within our adopted HI detection limit, the curves of H3, H4, and H5, stay flat at about 100 km/s, just above 80 km/s, and just below 80 km/s, respectively.  The velocities are consistent with their masses, and although we do not make a detailed comparison the curves are similar to observed rotation curves -- in particular, they do not posses large bulges with centrally peaked rotation curves, as seen in many simulations without effective feedback.  H1 has a higher rotational velocity (not surprising given its larger mass) and shows some mild features at 1 and 5 kpc.  On the other hand, the curve of H2 looks quite different: it is highly peaked, and falls sharply at about 4 kpc (again because of partial CR pressure support).  As can be seen from the edge-on plot of H2, this dwarf galaxy does have a dense core, larger than the other dwarfs. 

\begin{figure}
\centering
  \includegraphics[scale=0.45]{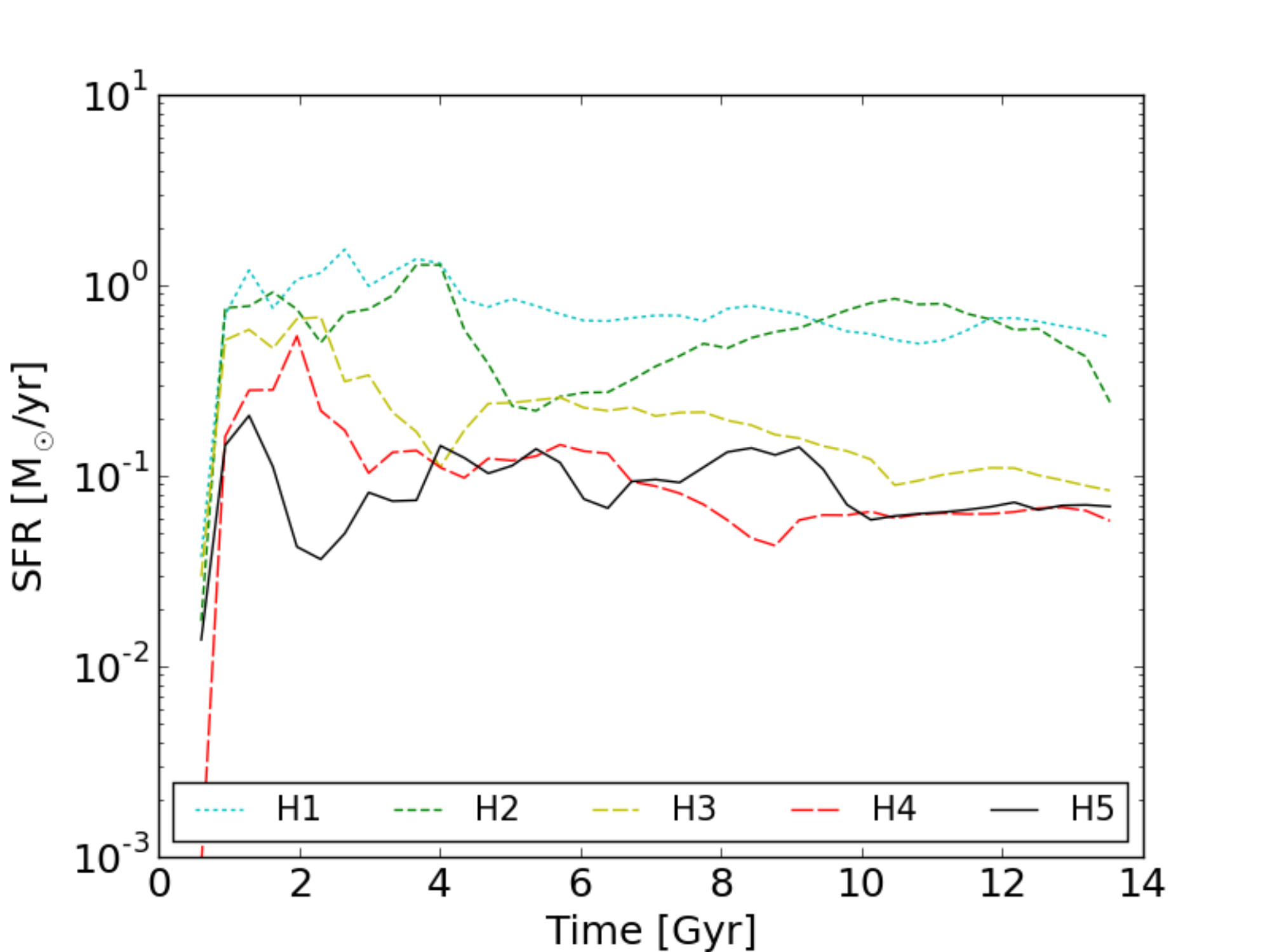}
  \caption{Star formation histories of our simulated halos H1 to H5 (all with fixed CR diffusion coefficient).}
  \label{mass_sfr}
\end{figure}

Figure~\ref{mass_sfr} shows the star formation histories of these same halos.  There is a clear progression of increasing SFR with increasing mass.  Despite some bumps and wiggles, the overall histories are remarkably flat, with each profile showing typically a factor of 2-3 variation over 12-13 Gyr.  There is a general trend for slowly falling star formation rate with time, particularly evident in H3.  This occurs because of the relatively tight self-regulation -- feedback acts to decrease the gas density, slowing star formation.


\subsection{Comparison with Observed Scaling Relations} \label{comparison_observation}

Dwarf galaxies are less efficient in star formation than normal galaxies, and the baryon fraction of galaxies decreases as the total mass goes down \citep[e.g.,][]{2010ApJ...708L..14M}.
In this section, we compare our simulation results with a range of observational probes, including the baryonic Tully-Fisher relation, as well as observed gas-to-star ratios and inferred stellar mass-to-halo mass relations.

\begin{figure}
\centering
\includegraphics[scale=0.45]{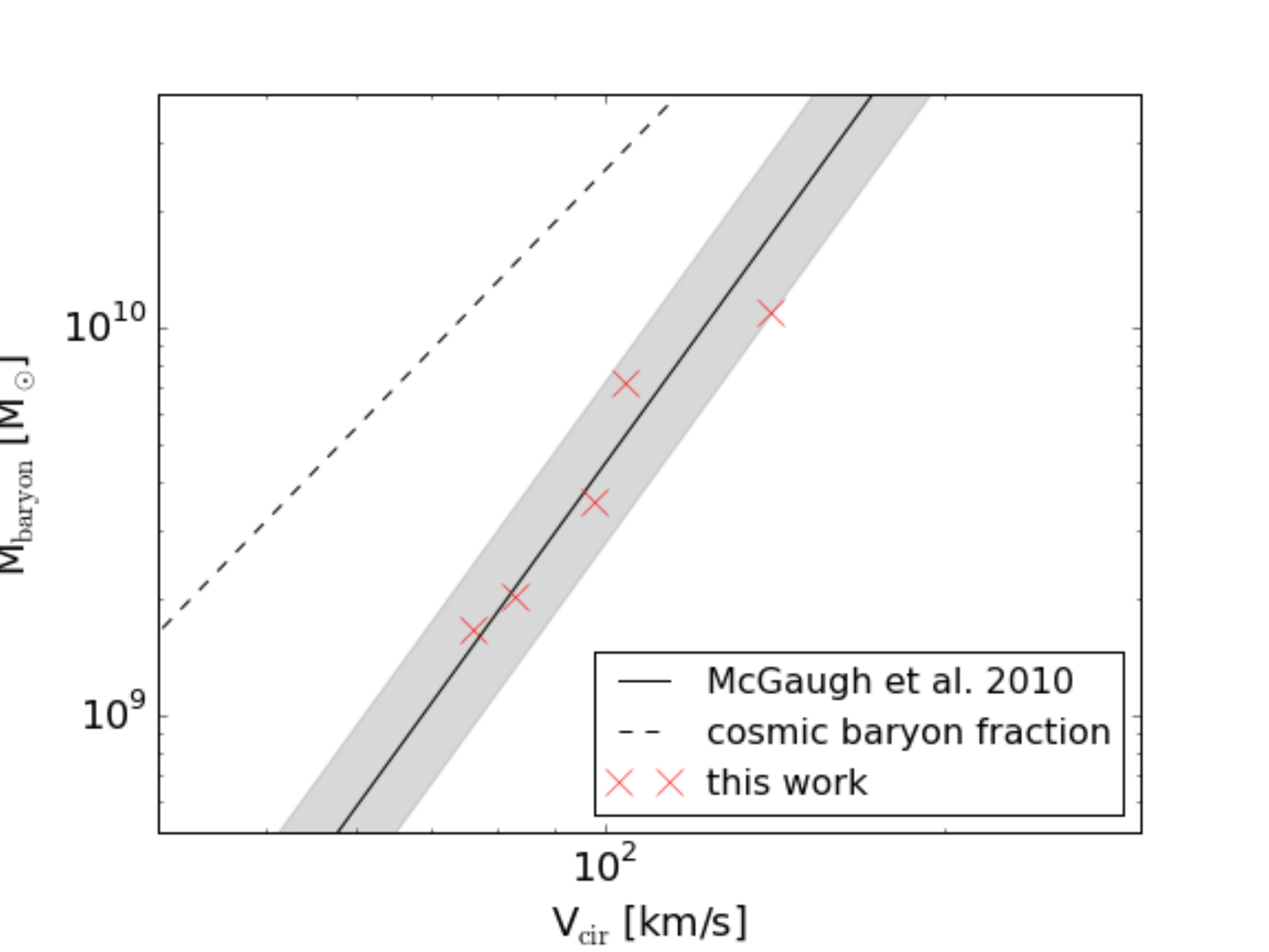}
\caption{Comparison between the five dwarf simulations (H1-H5) run with our standard diffusion coefficient and the baryonic Tully-Fisher relation from observations. The solid line is the fit from \citet{2010ApJ...708L..14M}. The red crosses are simulation data. The grey area shows the region of one sigma error in $M_b$.  The dashed line is taken from \citet{2009Komatsu} and indicates the relation between the halo circular velocity and the cosmic baryon fraction of the halo.}
\label{btf_overlay}
\end{figure}

We begin with the observed relation between the disk mass (the combined stellar and gas masses) and the flat part of the rotational velocity of the disk -- the baryonic Tully-Fisher relation.  Figure~\ref{btf_overlay} shows the simulation results from varying halo mass with fixed CR physics overlaid on the fitted line of the Baryonic Tully-Fisher relation from \citet{2010ApJ...708L..14M}.  They provide a fit for disk galaxies in the range of $20$ to $350$ km/s given by  M$_{b}  = x\ \mathrm{log}\ {V}_{c} + {A}$ with $x = 4.0$ and $A =1.65$.
In Figure~\ref{btf_overlay}, the solid line represents their best fit, and the grey area represents one standard deviation of $M_b$.  We estimated the scatter in the observed $M_b$ as the average for spiral and gas disk galaxies in Table 2 of \citet{2010ApJ...708L..14M}.

For the simulations, we determined the baryonic mass as the sum of the stellar and gas masses, with the gas mass calculated as $1/0.74$ times the HI mass.   Although the code evolves the HI density, it is underestimated in dense gas due to the lack of self-shielding.  To correct for this, the HI density is estimated as follows: when the total H number density $n_{\rm H} > 2\times10^{-2}$ cm$^{-3}$,  we assume that the neutral fraction is nearly unity and take $n_{\rm HI} = n_{\rm H}$; otherwise we assume the HI density is zero.  This is approximately consistent with radiative transfer calculations \citep{2013MNRAS.431.2261R} that take into account the metagalactic radiation background (a local radiation background would decrease the neutral gas fraction even lower, and so our bounds here are conservative). 

Both the stellar mass and HI mass are calculated within the same HI column density limit we used to determine the radial extent for the rotation curves ($10^{20}$ cm$^{-2}$) and extend above and below the plane by 2 kpc.  The circular velocity is calculated as the average rotation velocity of the gas at a radius of 2.2 scale lengths of the stellar surface density.  To determine the scale length, we fit the stellar surface density curve with a straight line from 2 kpc to 10 kpc.

Our simulation results all lie within the grey area and scatter around the observed relation.   Although we do not have a significantly large numerical sample to make a detailed comparison, these preliminary results appear to be in good agreement, indicating that the baryonic content of our galaxies is consistent with that observed.  We note that we cannot include the purely thermal simulation (H1-0) on this plot as the systems has no systematic rotation.

\begin{figure}
\centering
\includegraphics[scale=0.45]{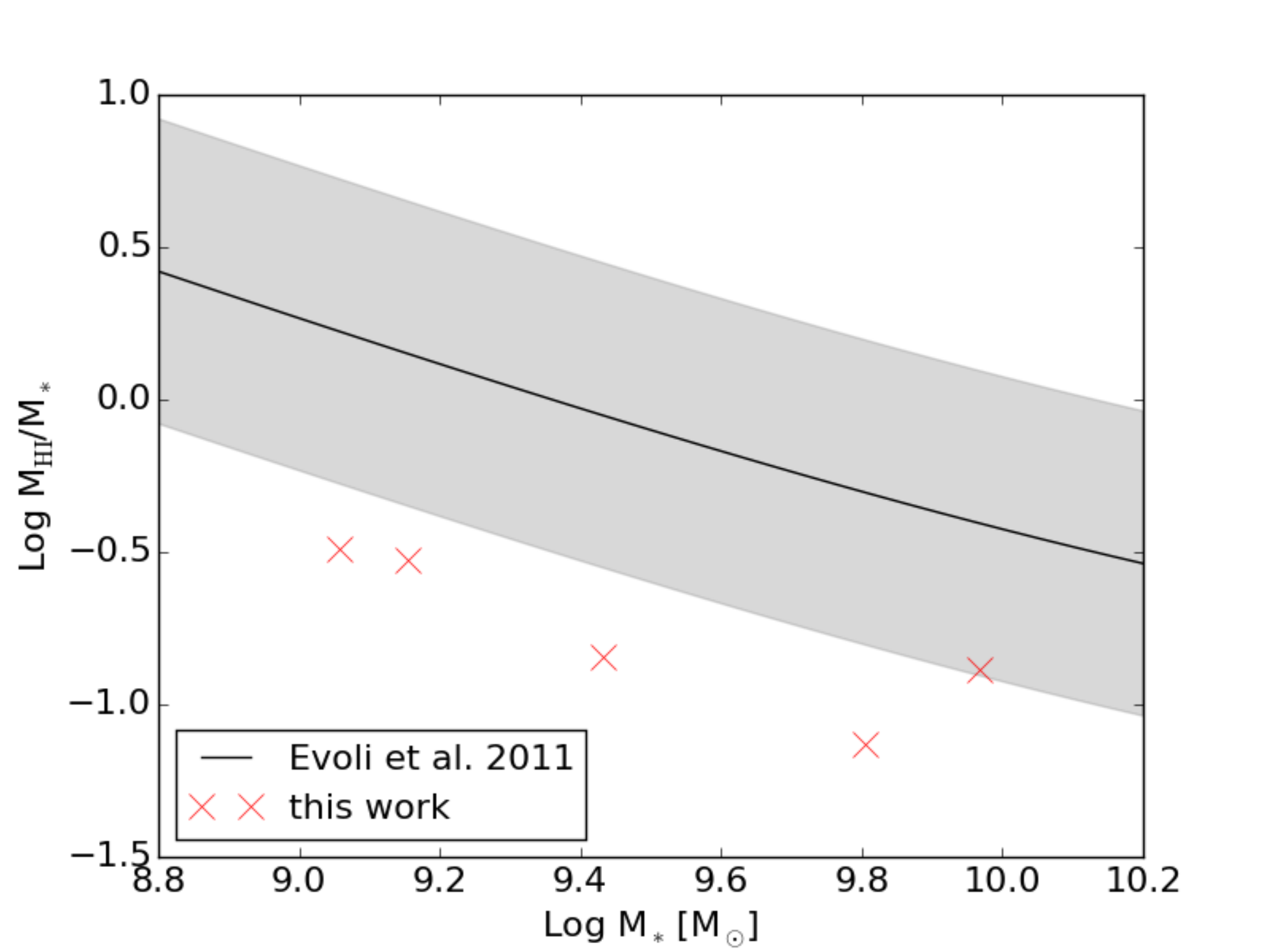}
\caption{Comparison between this work and the relation of HI mass and stellar mass inferred from observations. The solid line is derived in  \citet{2011ApJ...743...45E}. The grey area shows an estimated scatter of +/- 0.5 log around the observed relation in the vertical direction.  The red crosses are the simulation results. }
\label{hi_star}
\end{figure}

The baryonic Tully Fisher relation compares the galaxies' baryonic mass with a measure of its total mass.  We would also like to compare the two main components of baryons: neutral hydrogen and stars.  Figure~\ref{hi_star} shows the ratio of the HI to stellar components as a function of the stellar mass, all quantities which are directly observable.  The observed relation found in \citet{2011ApJ...743...45E} (their Eq. 5) can be approximated as:
\begin{equation}
\begin{split}
\frac{M_\mathrm{HI}}{3.36 \times 10^9 \rm{M}_\odot} = & 
\left(\frac{M_\star}{3.3 \times 10^{10} \rm{M}_\odot}\right)^{0.19} 
\\ & \times 
\left[1+\left(\frac{M_\star}{3.3 \times 10^{10} \rm{M}_\odot}\right)^{0.76}\right]
\end{split}
\end{equation}
In Figure~\ref{hi_star}, the solid line shows this relation.  We also estimate the observed scatter around this relation by taking a variation of +/- 0.5  log M$_\mathrm{HI}$/M$_\star$, based on a visual inspection of Figure~1 of \citet{2011ApJ...743...45E}; this is shown as a grey area in our Figure~\ref{hi_star}.  The ratio from the simulations is also plotted, again using the stellar and neutral hydrogen masses within our adopted HI surface density limit, and +/- 2 kpc in vertical extent.

As can be seen from Figure~\ref{hi_star}, all of our simulation results are below the observed relation, indeed most lie below the estimated scatter in the observations (the grey band).   This demonstrates that our simulation produces consistently lower HI fraction than observed.  There is some indication that the simulated relation shows a rising gas fraction with decreasing stellar mass; however, the sample size is too small to be sure.

This mismatch in the gas to star ratio, by approximately a factor of five, when combined with the agreement in the baryonic Tully-Fisher relation (which measures the total baryonic content), implies that our galaxies contain the correct amount of baryons, but are overly efficient in their conversion of gas to stars.  One alternative way to examine just the stellar content of our systems is based on the dark matter abundance matching measurements in \citet{2011ApJ...743...45E}, which provides a relation between the stellar mass and halo mass under the assumption that there is a monotonic relation between halo mass and stellar mass.  Indeed, if we compare our results to their inferred M$_*$-M$_{\rm halo}$ relation, we find that our $M_*$ values are systematically too large by factors of 2-5, consistent with the idea that the division between stellar and gas mass is incorrect in these simulations.  Finally, we note that these relations also imply that the simulations disagree with the observed Kenicutt-Schmidt relation between surface gas density and star formation surface density.

\subsection{Dark Matter Profiles}

There are a number of observational indications that dwarf galaxies do not contain the cuspy dark-matter profiles predicted in dark-matter only simulations \citep[e.g.,][]{2008deBlok, 2011Walker, 2011Boylan-Kolchin, 2012Boylan-Kolchin, 2012Salucci, 2014Klypin}.  A number of simulations with strong feedback have found that it is possible for the feedback to decrease the dark matter density in the center of halos \citep[e.g.,][]{1996Navarro, 2010Governato, 2012Pontzen, 2014Ogiya}.

\begin{figure}
\centering
\includegraphics[scale=0.45]{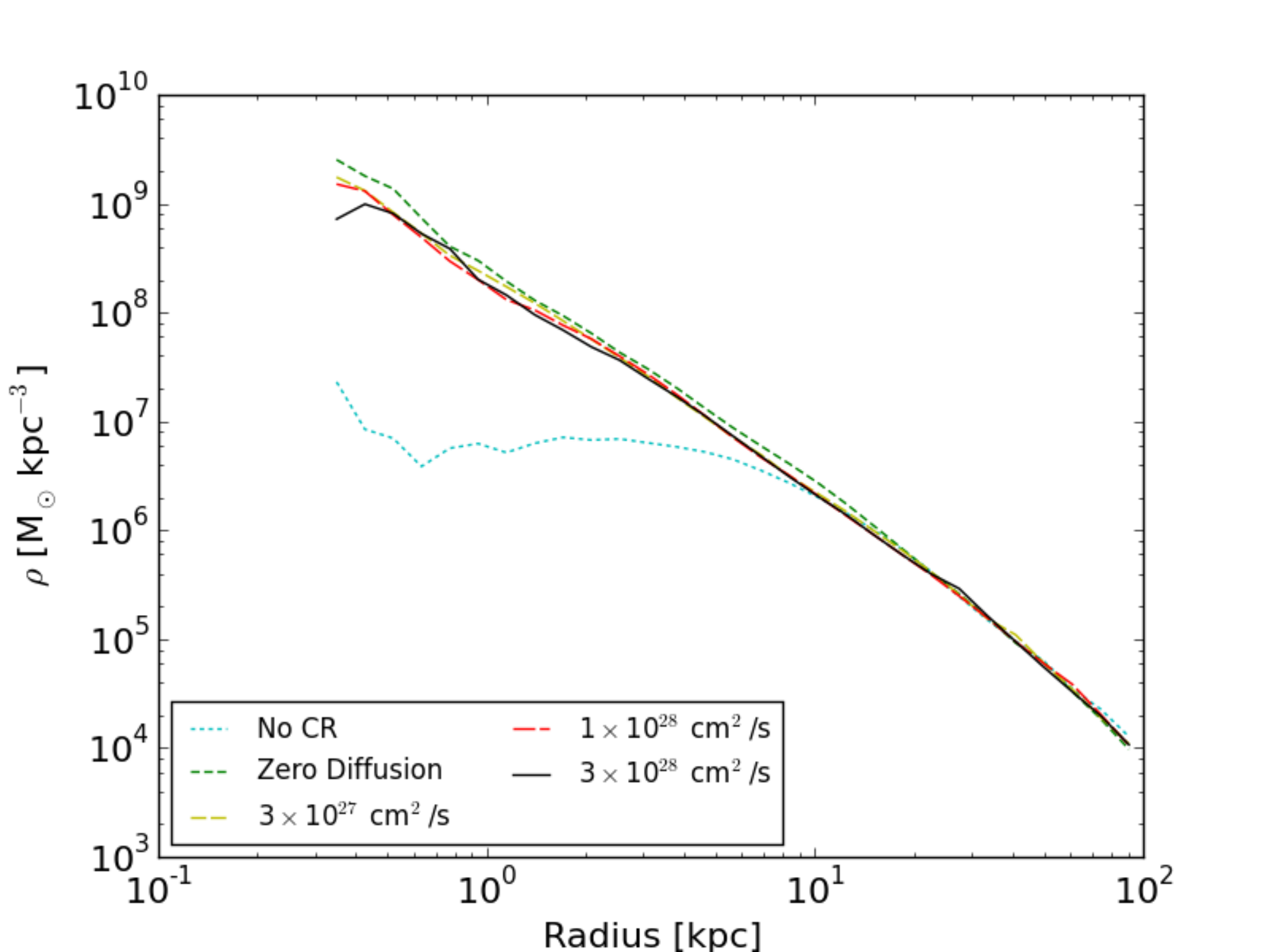}
\caption{Dark matter profiles of simulations with different CR physics.}
\label{cr_dm}
\end{figure}

Therefore, we examine our simulation to see the impact of thermal vs. CR feedback on the core properties of the systems.   Figure~\ref{cr_dm} shows the dark matter density profile of our most massive halo in the simulations with different CR physics.  Interestingly, all of the CR runs show very similar results, indicating that although they produce different distributions for the gas and stars, the presence of CR feedback leads to similar dark matter core properties.  In all cases, the profile is quite steep, with an inner density profile between $r^{-1}$ and $r^{-2}$ (similar to dark-matter only runs).  On the other hand, the run without CR diffusion shows a clear core in the dark matter with a core radius of several kpc.

The cored dark matter profile in the no-CR case might be due to the large and rapid motions of the gas, which can disturb the dark matter distribution \citep{2012Pontzen} and is consistent with the observed asymmetric gas distribution and the bursty nature of the feedback in the no-CR run.

\section{Discussion}
\label{sec:disc}

We have performed a preliminary investigation into the impact of CRs on the formation and evolution of dwarf galaxies with rotation velocities in the 70-140 km/s range using cosmological simulations.  This complements similar work \citep{2014SBH} which looked at a higher mass galaxy (with a rotation velocity above 200 km/s).  As in that work, we found that the inclusion of CR feedback had a strong impact on the gas and stellar properties.   In both cases, CR feedback led to robust outflows and rotationally supported disks with relatively flat rotation curves, but {\it only} if CR diffusion was included.  A comparison to observations showed good agreement with the baryonic Tully-Fisher relation, indicating that the total disk mass was reasonable.  This mass, converted into a fraction of the total halo mass, is well below the cosmic value, indicating that feedback was important in generating mass-loaded winds.  On the other hand, a comparison of the gas content of the simulated galaxies against observations showed that too much gas had been converted to stars.  In addition, the gas depletion times of our galaxies (computed as $M_{\rm gas}/\dot{M}_{\rm SFR}$), are a few Gyr, well below observed values for low-mass isolated dwarfs \citep{Huang2012}.

Our star formation and feedback model is approximate, with a number of free parameters required to describe physics that is not fully modeled.  In section~\ref{sec:vary_cr}, we explored the impact of some of these for halo H1.  Remarkably, we find that the CR diffusion rate has a relatively small impact on the stellar mass produced (see Table~\ref{diff_cr}), or the gas mass.  The exception was the run with $\kappa = 0$, which produced a substantially higher stellar mass and did not produce a disk (this combination does not generate winds).  

We also carried out an additional run of halo H1 with the standard diffusion rate (and other parameters), but a reduced CR energy efficiency (10\% instead of 30\%) and find that relative to the fiducial run, it produces 50\% more stars.  The resulting disk rotates marginally faster such that the system stays close to the observed baryonic Tully-Fisher relation.  However, the HI to gas mass ratio is even lower the fiducial run, resulting in an an even stronger disagreement with the observed relation.  In principle, a higher CR efficiency might bring this into agreement with observations, but that would disagree with local observational and theoretical work on CR acceleration efficiencies.

Another parameter which can be varied is the star formation efficiency per free-fall time, which we took to be 1\%.  We carried out another simulation much like H1 but with a star formation efficiency four times lower (with $\kappa = 3 \times10^{28}$ cm$^2$ s$^{-1}$; this run was not included in figures and tables above).  This had almost no effect on the resulting stellar mass or gas mass, consistent with the idea that the disk is in a state of self-regulation -- indeed, we verified visually that the lower efficiency resulted in a higher density gas disk and hence essentially the same net star formation rate at $z=0$ (since the star formation rate increases as $\rho^{1.5}$).

There are, of course, other parameters to vary, such as the IMF, the fraction of energy in SN, as well as other uncertainties in the subgrid model.  In addition, there is some indication from H2 and the high-mass halo explored in \citet{2014SBH} that an additional form of feedback is required in high-density regions -- possibly radiative heating or pressure.

Finally, we briefly discuss our results in comparison to \citep{2012MNRAS.423.2374U}, who recently carried out simulations of dwarf galaxies with cosmic rays.  The implementation differs somewhat in that they adopted a model in which CRs stream along magnetic field lines and heat the gas (rather than isotropically diffuse, as in our model).  Although they do not simulate cosmological volumes, their halo mass which most closely compares to ours produces approximately similar results: robust cosmic-ray driven outflows with a modest mass-loading factor.  Although we are unable to compute a mass-loading factor, the large stellar mass to halo mass we find implies a relatively low mass-loading factor.

\subsection{Feedback efficiency of the purely thermal run}
\label{sec:feedback_eff}

\begin{figure}
\begin{center}
\includegraphics[width=10cm]{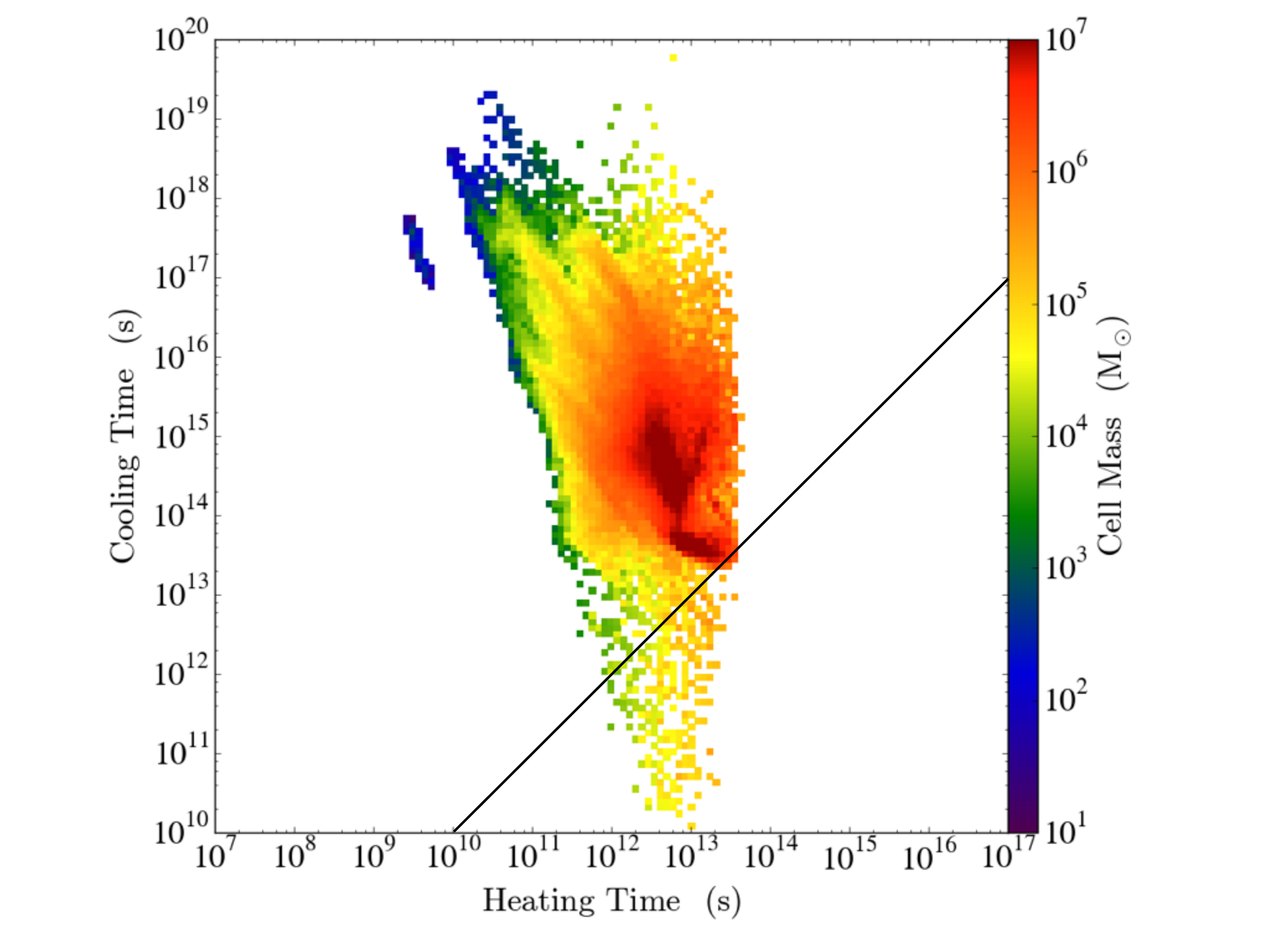}
\end{center}
\caption{The distribution of SN heating and radiative cooling times for all gas particles within 30 kpc of the center of our thermal-only simulation (H1-0) at $z=0$.  The black line indicates equality; since nearly all gas cells lie above this line, we see that pure thermal heating operates more rapidly than cooling.  See text for the definition of the SN heating time.}
\label{fig:crno_coolheat}
\end{figure}

Finally, we turn to a discussion of the thermal-only feedback run (H1-0).  Interestingly, the feedback in this case was actually {\it more} effective than the CR+thermal feedback, in contrast to the result for the higher mass halo with the same code and CR parameters discussed in \citet{2014SBH}.    One possibility is that the purely thermal feedback is more efficient than in more massive galaxies because the shallower potential well results in low gas densities, lowering the radiative cooling rate and permitting blast waves from SN feedback to accelerate the gas.  

We explore the relative importance of cooling and heating in Figure~\ref{fig:crno_coolheat}, which shows a two-dimensional distribution of radiative cooling times and a measure of the possible supernova heating rate for all gas within 30 kpc of the center of this halo as $z=0$.   We compute the SN heating rate in a way similar to that in \citet{2013MNRAS.432.1989S}, where we estimate the timescale required for our thermal energy injection mechanism to heat a cell to $T_6 = 10^6 K$ (the precise temperature selected does not affect our conclusions):
\begin{equation}
t_{\rm heat} = \frac{3}{2} \frac{M_{\rm cell} k T_6}{\mu m_H} \frac{t_f}{e_{\rm SN} M_* c^2}
\end{equation}
where $M_{\rm cell}$ is the gas mass cell in the cell, $\mu \approx 1$, $M_* = 10^5$ M$_\odot$, our typical stellar particle mass, $t_f = 10$ Myr, the timescale over which a star particle returns its supernova energy to the ISM in our model, and $e_{\rm SN} = 3 \times 10^{-6}$, as defined in the methods section.  We stress that not all cells are being heated at this rate, but it is the rate that a single star particle would typically heat its local gas and so is a good measure of the timescale over which feedback heating acts.  

From this figure, we see that nearly all the gas in the simulation lies above the black line indicating equality.  This demonstrates that the gas densities are sufficiently low that pure thermal feedback is efficient in heating the gas and driving outflows -- we focus on density because the cooling rate is proportional to the density squared.   A close examination of this figure indicates that a substantial amount of gas lies close to the line of equality (in fact, this gas is in the disk), and so we expect that the larger potential depths of higher mass halos will push gas beyond the line into a regime where purely thermal supernovae feedback cannot heat the gas.  This is consistent with the finding that our dwarf galaxy simulations do drive outflows while higher mass (e.g. Milky Way mass) systems do not.  Note that this statement depends on details of the simulations (e.g. see definition of $t_{\rm heat}$) and so may be a reflection of numerical models rather than physical quantities.  

This demonstrates that purely thermal feedback is sufficient to drive feedback but it does not answer the question why adding cosmic rays decreases the bursty outflows, as seen in Figure~\ref{cr_edge}.  The answer appears to be that cosmic rays actually increase the pressure in the diffuse circumgalactic gas surrounding galaxies.   The purely thermal run has a (thermal only) CGM pressure at a radius of 10 kpc of approximately $10^{-14}$ erg cm$^{-3}$, while even the CR run with the highest diffusion coefficient (H1), has a (mostly CR) pressure of nearly $10^{-13}$ erg cm$^{-3}$.  This larger pressure resists the impulsive outflows driven by bursts of supernovae in the disk (although it does not appear to eliminate the slower gas outflows driven by the CR pressure gradient).

Finally, we note a consequence of the effective thermal only feedback, which is that it resulted in bursty star formation at a low-level; and was so effective that the resulting stellar system was ``hot" -- that is, it was supported by velocity dispersion rather than rotation (an issue that has been raised in the context of other simulations -- see the discussion in \citet{2015Wheeler}.


\section{Summary} \label{sec:discuss_sum}

We simulate a series of dwarf galaxies with zoom-in cosmological simulations, and study the influence of CR feedback.
First we compare the outcome by varying the CR physics in different runs of the same galaxy halo.
The different CR physics include feedback without CR, CR feedback with zero diffusion, CR feedback with $\kappa_\mathrm{CR}$ equal to $3 \times 10^{27}$, $1 \times 10^{28}$, and $3 \times 10^{28}$ cm$^2$ s$^{-1}$. 
Then we fixed the CR diffusion at $\kappa_\mathrm{CR} = 3 \times 10^{28}$ cm$^2$ s$^{-1}$ and simulated five dwarf galaxies with different masses, ranging from 8 to 30 $\times 10^{10}$ M$_\odot$.
We summarize the results of this work as follows:
\begin{enumerate}
\item Adding CRs and some realistic level of diffusion consistently produced thin, extended disk galaxies with nearly flat rotation curves, in contrast to a case with purely thermal feedback, which produced a hot stellar systems.  The star formation rate in the CR systems was relatively constant, compared to the lower (and burstier) SFR in the purely thermal feedback model.
\item Simulations of our five dwarf galaxies with different masses but the same CR model matched well with the observed baryonic Tully-Fisher relation.  However, the five galaxies' neutral hydrogen fractions were always lower than what is observed, by a factor of about five and the stellar content was larger than observed (but the total baryonic content matched observations).
\item Simulations with CR feedback produce cuspy dark matter profiles, while our purely thermal feedback case resulted in a cored dark matter profile.
\end{enumerate}

These results indicate that the impact of cosmic rays in dwarf galaxies is potentially quite important, a conclusion which is in general agreement with previous work \citep[e.g.,][] {2008Jubelgas}.  In detail, the results are somewhat dependent on the physical model, but broadly speaking CRs with diffusion produce dwarf galaxies which are rotationally supported and have relatively flat rotation curves.  This result appears to be quite robust (and occurs even though 70\% of the feedback energy is in the form of thermal energy in these models) and is probably due to the smooth pressure provided by CRs (along with the outflows they drive).  The baryonic content of the dwarf systems with CRs is in broad agreement with observations.

On the other hand, there are two areas of disagreement with observations in these models: (a) the detailed census of baryons in each system (stars vs. gas) is much too heavily weighted to the stellar side, and (b) the presence of dark matter cores may be a problem (although CR pressure may systematically support the gas at small radius and so bias the observed measures of dark matter on small scales).  Clearly, more work is required to explore and refine these results.  For example, it would be useful to explore more realistic models including magnetic fields, anisotropic CR diffusion, and the impact of field-CR streaming instabilities.


\section*{Acknowledgments}
We thank the anonymous referee for a useful report which improved the clarity of the paper.
We would like to thank Cameron Hummels, Mordecai-Mark Mac Low and Mary Putman for useful discussion related to this work. We acknowledge financial support from NSF grants AST-1312888, and NASA grants NNX12AH41G and NNX15AB20G, as well as computational resources from NSF XSEDE, and Columbia University's Yeti cluster.  Computations described in this work were performed using
the publicly-available {\sc Enzo} code (http://enzo-project.org), which
is the product of a collaborative effort of many independent scientists
from numerous institutions around the world. Their commitment
to open science has helped make this work possible.

\bibliography{ms}
\end{document}